# The affinity-efficacy problem: an essential part of pharmacology education

James P Higham[1], David Colquhoun[2]

[1]Research Associate, Department of Pharmacology, University of Cambridge, Tennis Court Road, Cambridge, CB2 1PD

[2]Professor Emeritus, Neuroscience, Physiology and Pharmacology, University College, London, WC1E 6BT

Correspondence should be addressed to: DC, d.colquhoun@ucl.ac.uk, or JPH, jph87@cam.ac.uk.

Key words: agonism, ligand binding, receptor theory, affinity, efficacy

## Abstract

A fundamental mistake in receptor theory has led to an enduring misunderstanding of how to estimate the affinity and efficacy of an agonist. These properties are inextricably linked and cannot be easily separated in any case where the binding of a ligand induces a conformation change in its receptor. Consequently, binding curves and concentration-response relationships for receptor agonists have no straightforward interpretation. This problem – the affinity-efficacy problem – remains overlooked and misunderstood despite it being recognised in 1987. To avoid the further propagation of this misunderstanding, we propose that the affinity-efficacy problem should be included in the core curricula for pharmacology undergraduates proposed by the British Pharmacological Society and IUPHAR.

## 1. Introduction

In 1956, RP Stephenson proposed a modification of receptor theory which has had a lasting influence [1]. He said, in addition to the affinity of a drug for a receptor, an extra parameter was needed to describe the action of an agonist. This extra parameter he called *efficacy*, which he defined as a measure of the agonist to produce a response once it became bound. This idea provided a unified understanding of agonists, partial agonists, and antagonists. However, a mistake in his framing of the problem resulted in a fundamental misunderstanding of how to estimate values of affinity and efficacy, a misunderstanding which still pervades pharmacology today. The resolution of this mistake, which was uncovered in 1987 [2], demonstrated, somewhat counterintuitively, that agonist binding depends on both its affinity *and* its efficacy [3]; a consideration which is all-too-often overlooked in the teaching of pharmacology and the pharmacological literature. This review is intended to explain the problem, commonly referred to as the affinity-efficacy, or binding-gating, problem.

## 2. A simple agonist mechanism

The problem in question in best illustrated by first considering the framework proposed by Stephenson [1]. He proposed that the fractional receptor occupancy ($p$) for an agonist (A) with a dissociation equilibrium constant, $K_A$, (defined as the ratio of the rate constants for dissociation [$k_{off}$] and association [$k_{on}$] of the ligand), is given by the Langmuir equation

$$p = \frac{[\mathrm{A}]}{[\mathrm{A}] + K_\mathrm{A}} \tag{1}$$

Stephenson postulated that the response to the agonist could be written as a function of the product of the occupancy and the efficacy of the agonist. The function was unknown but assumed to be the same for all agonists. Efficacy was an empirical number between zero and infinity; no physical interpretation of efficacy was proposed. Crucially, Stephenson assumed that affinity and efficacy were separable properties, an assumption that proved to be fatal for his approach.

The Langmuir equation, which was first derived by Hill in 1909 [4], describes the binding of a ligand to identical, independent binding sites at equilibrium. That would be sufficient to describe the binding of a competitive antagonist to a receptor, given that the antagonist does not produce a conformational change in the receptor, i.e., only the vacant and bound inactive states of the receptor are present [5]. But it cannot be expected to describe the action of an agonist, which must produce a conformational change in the receptor to produce a response. Neither can the Langmuir equation account for the phenomenon of partial agonism. The maximum occupancy at high ligand concentration is always one, but not all agonists elicit the same maximum response when all receptors are occupied.

The simplest possible mechanism that overcomes these problems was proposed a year after Stephenson's paper, by del Castillo and Katz (1957) [6]. Stephenson had hoped to solve the problem without postulating a mechanism for agonist action, whereas del Castillo and Katz proposed a simple mechanism. They added an extra state: after the agonist had bound to the receptor, it could isomerise to an active conformation. If we denote the agonist molecule as A, and the receptor as R (in its inactive conformation) or R* (in its active conformation), the del Castillo-Katz mechanism can be written thus

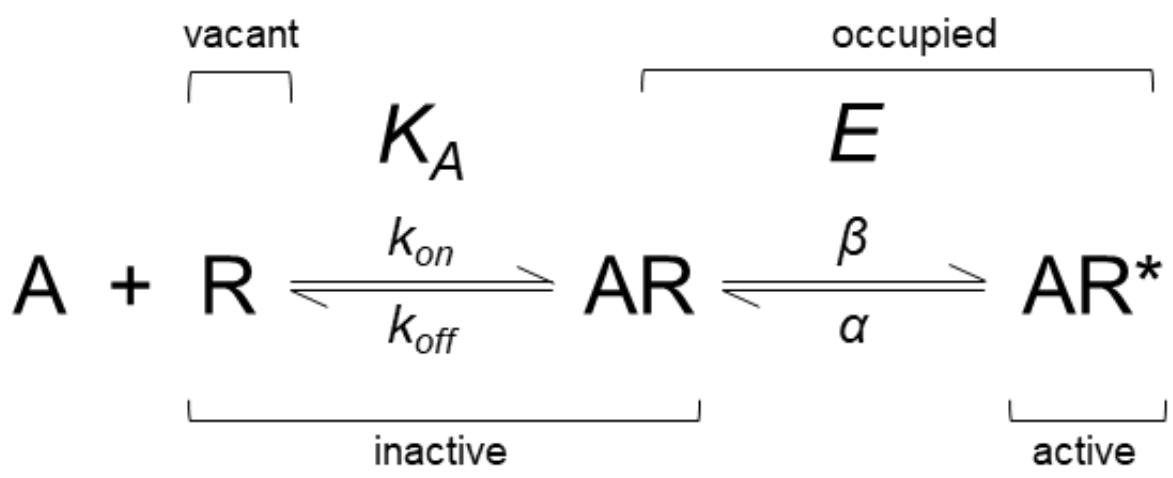

(2)

The equilibrium constant for the binding reaction is $K_A$, as before. The equilibrium constant for the isomerisation between the inactive and active conformations is denoted *E* because this is a measure of the agonist's efficacy. *E* is defined as the ratio of the rate constants for isomerisation to ($\beta$) and from ($\alpha$) the active state, AR*.

The mechanism in (2) is directly analogous with the Michaelis-Menten mechanism [7], which was published in 1913 and has been used as a simple description of the rate of action of enzymes ever since. In fact, the mechanism in (2) is even simpler than the Michaelis-Menten case, in which the last step of the reaction is usually assumed to be irreversible, so it can't attain true equilibrium, but only a quasi-steady state.

In (2), both of the states AR and AR* are bound by agonist, so to calculate the receptor occupancy that would be found in an equilibrium agonist binding experiment, it's necessary to add the occupancies of these two states, $p_{AR}$ and $p_{AR^*}$. Application of the law of mass action to (2) shows that

$$p_{AR} = \frac{[\mathrm{A}]p_{\mathrm{R}}}{K_{\mathrm{A}}}$$

$$p_{\mathrm{AR}^*} = E p_{\mathrm{AR}}$$

where $p_R$ is the fractional occupancy of the vacant state. Consequently, the total fraction of receptors bound by agonist ($p_{bound}$) is

$$p_{\mathrm{bound}} = p_{\mathrm{AR}} + p_{\mathrm{AR}^*} = \frac{[\mathrm{A}]}{[\mathrm{A}] + K_{\mathrm{eff}}}$$

(3)

The derivation of this result has been omitted for brevity; it is explained in detail in two videos [8,9]. The result looks exactly like the Langmuir equation, eq. (1), except it contains a different equilibrium constant – not the affinity, $K_{\mathrm{A}}$, but rather an *effective* equilibrium constant, $K_{\mathrm{eff}}$, defined as

$$K_{\mathrm{eff}} \equiv \frac{K_{\mathrm{A}}}{1 + E}$$

(4)

The equilibrium constants for a single step in the reaction, like $K_{\mathrm{A}}$, are called microscopic constants, whereas constants that describe the net result of more than one step, like $K_{\mathrm{eff}}$, are called macroscopic constants.

This shows that the binding of an agonist to a receptor is dependent not only on its affinity but also on its efficacy. In Stephenson's framework, affinity and efficacy are not separable as he had assumed. This conclusion also follows, more generally, from thermodynamic considerations: the principle of reciprocity states that if binding affects the equilibrium between R and R*, which is the case for an agonist, then the reverse must also be true (e.g., [10,11]).

*Stephenson's mistake*

Stephenson postulated that the proportion of receptors which are occupied by an agonist at equilibrium depended only on $K_A$ as in eq. (1). In the 1980s, one of us (DC) asked Stephenson whether the word *occupancy* in this statement referred to that which would be measured in a ligand binding experiment. When Stephenson answered 'yes', it became obvious that he'd made a fundamental error. As shown in eq. (3) the amount of agonist bound depends on two different equilibrium constants; it depends on *both* the affinity *and* the efficacy. The agonist binding curve is predicted to have the same shape as a Langmuir isotherm, but the concentration for half-maximal binding, $K_{\mathrm{eff}}$, depends on both the affinity for the initial binding step, $K_{\mathrm{A}}$, and on efficacy, *E*, as shown in eq. (4).

It's clear from eq. (3) that the maximum occupancy is one. In other words, (virtually) all receptors become bound at a very high concentration of agonist. In the del Castillo-Katz mechanism, the response to the agonist is represented by the fraction of receptors in the active state, $p_{\mathrm{AR}^*}$. As the sum of the fractional occupancies of each state in (2) must equal one, the law of mass action implies that

$$p_{AR^*} = \frac{Ec_{\mathrm{A}}}{1 + c_{\mathrm{A}} + Ec_{\mathrm{A}}} \tag{5}$$

where we have expressed the concentration of the agonist as a multiple of its equilibrium dissociation constant, by defining the dimensionless variable

$$c_{\mathrm{A}} \equiv \frac{[\mathrm{A}]}{K_{\mathrm{A}}} \tag{6}$$

At a high concentration of agonist, when all receptors are occupied, the maximum fraction of receptors in the active state, $p_{\mathrm{AR}^*}^{\max}$, is given by the limit of eq. (5) as the free agonist concentration becomes very high. In this case, all receptors are occupied (either as AR or as AR*), so $p_{\mathrm{AR}^*}^{\max}$ depends only on efficacy and is given by

$$p_{\mathrm{AR}^*}^{\max} = \frac{E}{1 + E} \tag{7}$$

So, for an agonist with *E* = 1, the maximum response evoked by a saturating concentration of agonist will correspond to half of the receptors being in the active state (AR*, Figure 1A), e.g., half of the ion channels are open. The other half will be in the inactive but bound state (AR). In this case, the agonist will be obviously partial when compared with an agonist which evokes a larger maximum response. If *E* = 20, then the maximum response will correspond to 20/21 ≈ 95.2% of receptors in the active state. In most sorts of experiment, this would be indistinguishable from an agonist with *E* = 1000, which gives $p_{\mathrm{AR}^*}^{\max}$ = 1000/1001 ≈ 99.9% of receptors in the active state (Figure 1A). So, it's impossible to distinguish between agonists with an efficacy greater than 10 or so by measuring the maximum responses they elicit. Beyond this point, increases in efficacy merely shift the log concentration–response curve to the left, an effect which is indistinguishable from an increase in affinity [1].

As Stephenson pointed out, an agonist with $E = 1$ will be obviously partial only if the response being measured is the fraction of receptors in the active state, as in the case for ion channels. For responses that are limited by things other than receptor saturation, it's possible that a maximum response will be elicited by activation of a small fraction of receptors. In this case there are said to be *spare receptors.*

It is worth noting that for a ligand with $E = 0$, *i.e.*, a silent antagonist, then $K_{eff} = K_A$; the binding of an antagonist will not affect the equilibrium between the R and R* conformations of the receptor. In this case, the Langmuir equation provides a good description of ligand binding, wherein only the vacant and bound inactive states of the receptor are present. That is why the Schild method [5] for the estimation of the affinity of a competitive antagonist works even when the relation between the number of active receptors and the observed response is not known. Stephenson tried to apply similar null methods to agonists, but this proved to be impossible.

*Implications for structure-activity studies*

All this has profound implications for the interpretation of structure-activity relationships for receptors. In eq. (2), it is the equilibrium constant for binding, $K_A$, which gives information about the agonist's affinity for its binding site. The other equilibrium constant, $E$, tells you about the ability of the agonist to induce a conformation change in the receptor once it has become bound.

Both affinity and efficacy influence agonist binding, as shown in eq. (3), and, crucially, *the ability to change conformation can be influenced by mutations in any part of the receptor protein*. That's why changes in agonist binding don't necessarily tell you anything about the binding site. The same is true for structure-activity relationships for agonists. If a change in the chemical structure of an agonist leads to a change in agonist binding, it is not possible to deduce whether this is due to a change in agonist affinity or efficacy (from the binding experiment alone).

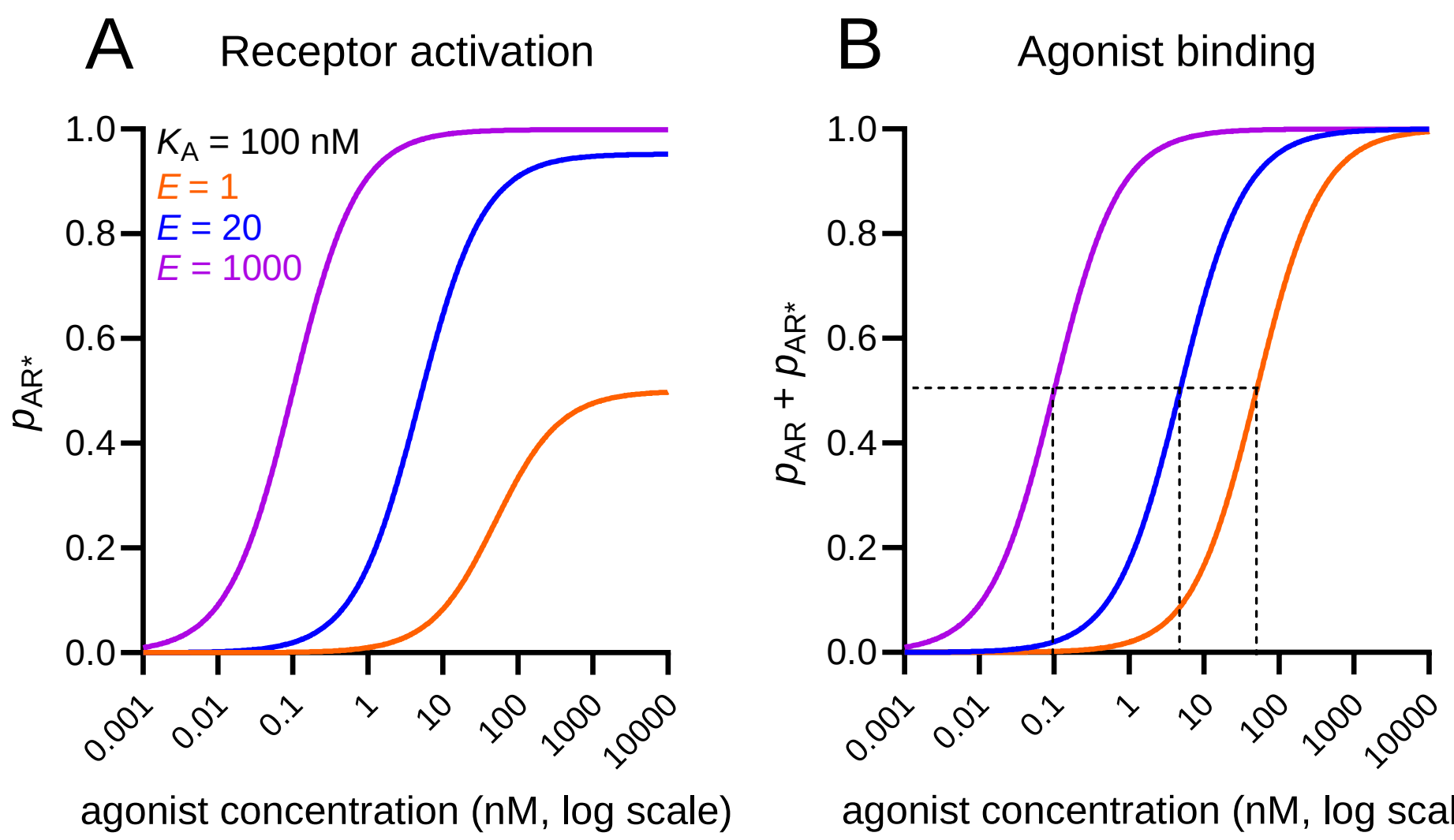

Figure 1 Illustration of the effect of changing efficacy for agonists with the same affinity. (A) The fraction of receptors occupying the active, AR*, state, calculated using eq. (5), for agonists which all have the same affinity ($K_A$ = 100 nM for all curves), but with differing efficacy ($E$ = 1, orange curve; $E$ = 20, blue curve; $E$ = 1000, purple curve). (B) Binding of an agonist, calculated from del Castillo-Katz mechanism, eq. (3). Left curve (purple), agonist with $K_A$ = 100 nM and $E$ = 1000. Middle curve (blue), agonist with the same affinity, $K_A$ = 100 nM, but reduced efficacy, $E$ = 20. Right curve (orange), agonist with the same affinity, $K_A$ = 100 nM, but further reduced efficacy, $E$ = 1. The grey dashed lines mark the concentrations at which half of the receptors are bound by agonist. These are $K_{eff}$ = 100/(1000+1) ≈ 0.1 nM (purple curve), 100/(20+1) ≈ 4.8

nM (blue curve) and 100/(1+1) = 50 nM (orange curve). The reduction in efficacy has reduced binding by a factor of 4.8/0.1 = 48-fold and 50/0.1 = 500-fold, respectively, with no change in affinity.

These ideas are illustrated in Figure 1B, which shows the binding of an agonist, as calculated from eq. (3) and (4). For each curve, the affinity is identical, $K_A$ = 100 nM. The leftmost (purple) curve is for an agonist with very high efficacy, $E$ = 1000. The middle curve (blue) is for an agonist with much reduced efficacy, $E$ = 20 (this could be a different agonist on the same receptor, or it could be the same agonist on a mutated receptor). The 48-fold reduction in macroscopic binding results entirely from reduced efficacy, so it doesn't necessarily tell us anything about the binding of the agonist to its binding site. The rightmost curve (orange) is for a low efficacy agonist ($E$ = 1), for which half-maximal binding is attained at a concentration of 50 nM. This represents a 500-fold reduction in macroscopic binding, again with no change in $K_A$.

Agonist binding experiments are, therefore, of limited use in the elucidation of agonist affinity, or of the location of the agonist binding site. One should be wary when reporting the results of agonist binding experiments as they provide only the effective equilibrium constant, $K_{eff}$, not the dissociation constant, $K_A$. Concentration-response measurements provide very similar information.

All this being said, if the effects are big enough, it may be possible to justify conclusions about the binding site from such studies, but they need to be buttressed by quantitative arguments, as, for example, in Anson *et al.* (1998) [12]. Measurements of the change in affinity of competitive antagonists caused by mutations in the putative binding site (measured by the Schild method or in ligand binding experiments) provide stronger evidence for the location of the binding site, insofar as the antagonist does not cause a conformational change in the receptor [12].

## 3. A more complex agonist mechanism

The del Castillo-Katz mechanism is too simple to describe any real receptor, but it is sufficient to show the mistake in Stephenson's framework. We will next consider a more realistic agonist mechanism, that for the interaction between acetylcholine (ACh) and related compounds and the muscle nicotinic acetylcholine receptor (nAChR) proposed in 1985 [13], shown in Figure 2. At the time, this was the best description of the action of ACh at the muscle nAChR, though it was later amended to include an intermediate inactive (shut) conformation which cast light on the mechanism of partial agonists [14].

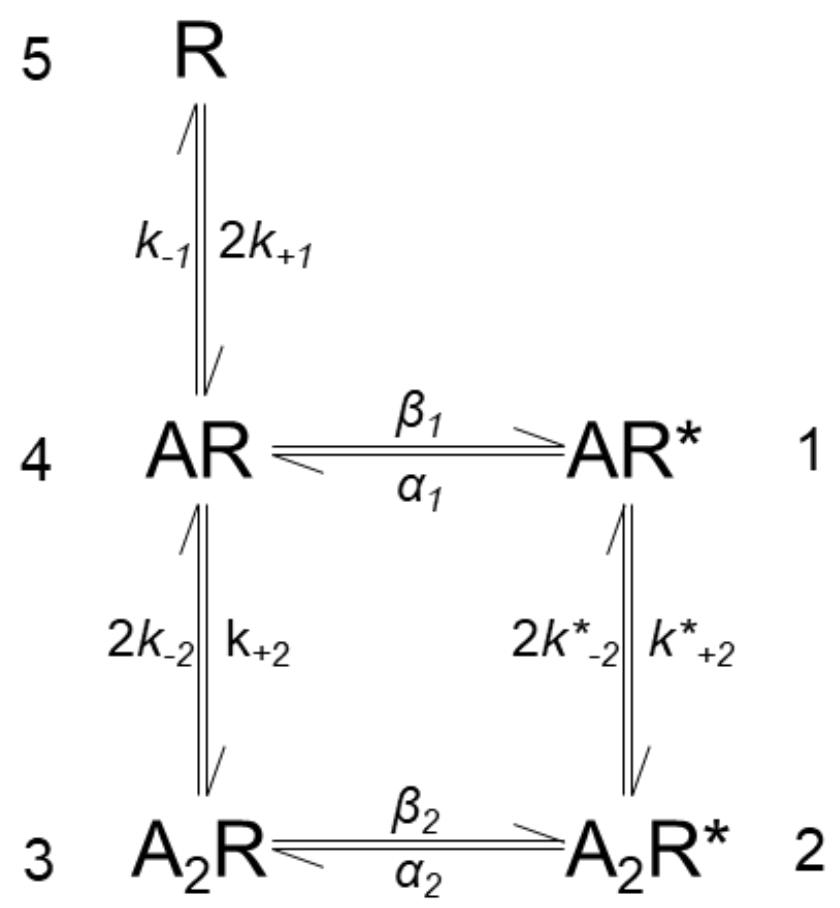

Figure 2 The mechanism for the nAChR proposed by Colquhoun and Sakmann (1985). Numbers adjacent to each state refer to their positions in the ***p*** and ***Q*** matrices (see text). Rate constants are given next to their respective transitions; some rate constants (e.g., the rate constant for the state 5 → 4, 3 → 4 and 2 → 1 transitions) are doubled because either one of two ligands may associate or dissociate during these transitions. This ensures that the rate constant refers to the binding site, rather than the receptor as a whole.

The mechanism proposed in 1985 (Figure 2) consists of three inactive states (channel is shut) and two active states (channel is open). While the binding of one agonist molecule can cause the opening of the channel, this reaction is quite unfavourable; the binding of two agonist molecules is far more likely to result in channel opening. To find the fractional occupancy of each state of the receptor, one could use the approach above for the del Castillo-Katz mechanism and apply the law of mass action to generate expressions for the occupancy of each state. While there is nothing wrong with this approach, it is somewhat cumbersome – and will become increasingly cumbersome for mechanisms with increasing numbers of states.

A more general way to solve for the fractional occupancies (for any mechanism) is to use matrix notation (section 3 in Colquhoun and Hawkes, 1995 [15]). Matrices are, in fact, the only way to deal with this sort of problem in a general way. If you are unfamiliar with them, try this video: *Matrix algebra in 45 minutes* [16].

First, a row vector containing the fractional occupancies, at time *t*, of each of the five states (***p***(*t*)), numbered as in Figure 2, is defined

$$\boldsymbol{p}(t) = [p_1(t) \quad p_2(t) \quad p_3(t) \quad p_4(t) \quad p_5(t)]$$

Any reaction mechanism can be defined by its transition rate matrix, which is usually denoted the ***Q*** matrix. This contains the rates for the transitions between each state. The off-diagonal element, $q_{ij}$, in the $i^{\text{th}}$ row and $j^{\text{th}}$ column is the rate of the transition between states *i* and *j*. If there is no direct connection between two states, the transition rate between them is zero. The diagonal elements are defined such that the sum of each row in the ***Q*** matrix is zero (and, consequently, its determinant is also zero). For example, for the mechanism in Figure 2, the ***Q*** matrix is:

$$\boldsymbol{Q} = \begin{bmatrix} -([\mathrm{A}]k_{+2}^{*} + \alpha_1) & [\mathrm{A}]k_{+2}^{*} & 0 & \alpha_1 & 0 \\ k_{-2}^{*} & -(k_{-2}^{*} + \alpha_2) & \alpha_2 & 0 & 0 \\ 0 & \beta_2 & -(2k_{-2} + \beta_2) & 2k_{-2} & 0 \\ \beta_1 & 0 & [\mathrm{A}]k_{+2} & -(\beta_1 + [\mathrm{A}]k_{+2} + k_{-1}) & k_{-1} \\ 0 & 0 & 0 & 2[\mathrm{A}]k_{+1} & -2[\mathrm{A}]k_{+1} \end{bmatrix}$$

Using this notation, it is possible to rewrite the set of five simultaneous differential equations needed to describe the rate of change in the occupancy of each state (omitted here for brevity) simply as

$$\frac{\mathrm{d}\boldsymbol{p}(t)}{\mathrm{d}t} = \boldsymbol{p(t)Q}$$

This equation is the same for any mechanism; all you need to do is to specify its ***Q*** matrix. At equilibrium (after infinite time), the occupancies don't change with time: d***p***(*t*)/d*t* = **0**, so

$$\boldsymbol{p}(\infty)\boldsymbol{Q} = \boldsymbol{0}$$

where $\boldsymbol{p}(\infty)$ is a row vector containing the fractional occupancies of each state at equilibrium. It is not immediately clear how to solve this for $\boldsymbol{p}(\infty)$, because the ***Q*** matrix is singular (its determinant is zero) and so it can't be inverted. Several ways exist to solve for $\boldsymbol{p}(\infty)$, but the most convenient for programming uses the augmented ***Q*** matrix [15]. This approach involves creating a new matrix (***S***) by augmenting the ***Q*** matrix with a unit column vector (***u***) which constrains the sum of the fractional occupancies to one. The ***S*** matrix can be written, in partitioned form, as

$$\boldsymbol{S} = [\boldsymbol{Q} \quad \boldsymbol{u}]$$

Hence, the transpose of ***S*** is

$$\boldsymbol{S}^{\mathrm{T}} = \begin{bmatrix} \boldsymbol{Q}^{\mathrm{T}} \\ \boldsymbol{u}^{\mathrm{T}} \end{bmatrix}$$

Post-multiplying $\boldsymbol{p}(\infty)$ by $\boldsymbol{SS}^{\mathrm{T}}$ gives

$$\boldsymbol{p}(\infty)\boldsymbol{SS}^{\mathrm{T}} = \boldsymbol{p}(\infty)\boldsymbol{QQ}^{\mathrm{T}} + \boldsymbol{p}(\infty)\boldsymbol{uu}^{\mathrm{T}}$$

As we are considering the system at equilibrium, $\boldsymbol{p}(\infty)\boldsymbol{Q} = 0$, so the first term on the right is zero. And $\boldsymbol{p}(\infty)\boldsymbol{u}$ is the sum of the fractional occupancies of all states at equilibrium which must equal one, so the second term is $\boldsymbol{u}^{\mathrm{T}}$. Therefore, this equation simplifies to

$$\boldsymbol{p}(\infty)\boldsymbol{SS}^{\mathrm{T}} = \boldsymbol{u}^{\mathrm{T}}$$

The determinant of $\boldsymbol{SS}^{\mathrm{T}}$ is non-zero, so it can be inverted, and post-multiplying both sides by $(\boldsymbol{SS}^{\mathrm{T}})^{-1}$ gives the equilibrium fractional occupancies for each state as

$$\boldsymbol{p}(\infty) = \boldsymbol{u}^{\mathrm{T}}(\boldsymbol{SS}^{\mathrm{T}})^{-1} \tag{8}$$

Numerical examples of these calculations are given in [9,17]. This approach enables one to simultaneously find the equilibrium occupancies for all states of a receptor for *any* given mechanism.

Notice that the ***Q*** matrix contains all of the rate constants in the mechanism. This means that the binding of an agonist – as well as the response to an agonist (and the time constants for approach to equilibrium) – depends on the *entire* reaction mechanism, including parameters describing the agonist's efficacy, and not just the affinity for the initial binding step. Eq. (8) is the general form of the "occupancy equation": it gives the occupancies of every state at equilibrium, for any reaction mechanism, while eq. (1) and eq. (3) are special cases pertaining only to specific reaction mechanisms.

## 4. The case of single ligand-gated ion channels

The ***Q*** matrix is at the heart of calculations of not only equilibrium properties, but also kinetics (the rate of approach to equilibrium), for both macroscopic and single molecule systems [15,17,18]. These can all be calculated from the ***Q*** matrix. It's possible to write a similarly general equation which describes the rate at which the occupancy of each state approaches its equilibrium value

$$\boldsymbol{p}(t) = \boldsymbol{p}(0)\exp(\boldsymbol{Q}t)$$

where $\boldsymbol{p}(t)$ is a row vector containing the occupancies for each state at any time, $t$, and $\boldsymbol{p}(0)$ is a row vector containing the occupancies for each state at time $t$ = 0. This simple-looking equation describes the kinetics for any mechanism (as long as ***Q*** is constant) but it is beyond the scope of this review - see [9,15,17,18].

In order to evaluate any of these equations, one needs numerical values for all of the rate constants in the reaction mechanism. The problem of estimating these values from experimental observations has, so far, been solved only for single ligand-gated ion channel recordings.

One problem in estimating the rate constants in a reaction mechanism resulted from the fact that very short channel openings or shuttings (shorter than ~20 µs) can't be resolved in experimental recordings. This means that observed openings are too long (because of missed short shuttings) and observed shut times are extended by missed short openings. In 1990, an exact solution to this problem was found by Hawkes and Jalali [19–21]. This allowed the calculation of the probability distributions of the *observed* open and shut times, given values for the transition rates in the ***Q*** matrix. And this, in turn, allowed the calculation of the probability (density) of observing the whole sequence of open and shut times in a recording, given putative values for the transition rates. This quantity is known as the likelihood of the set of putative transition rates. Their values are adjusted iteratively until the likelihood is maximised [22,23]. The maximum likelihood estimates of the transition rates, so found, can then be used to predict any other responses of the receptor, and its ability to bind the ligand.

Simulations have shown that these methods can provide good estimates of the rate constants, and hence of equilibrium constants, for mechanisms with up to 10 states and 14 free rate constants, under favourable conditions [24,25]. Thus, the problem of separating affinity and efficacy is solved.

For example, methods like these have allowed us to postulate that there is a short-lived shut conformation (the 'flipped', or 'primed', conformation) that lies between the resting conformation and the open conformation of channels in the nicotinic receptor superfamily [14,25,26]. This implies that three states exist in the presence of a saturating agonist concentration, as in Figure 3.

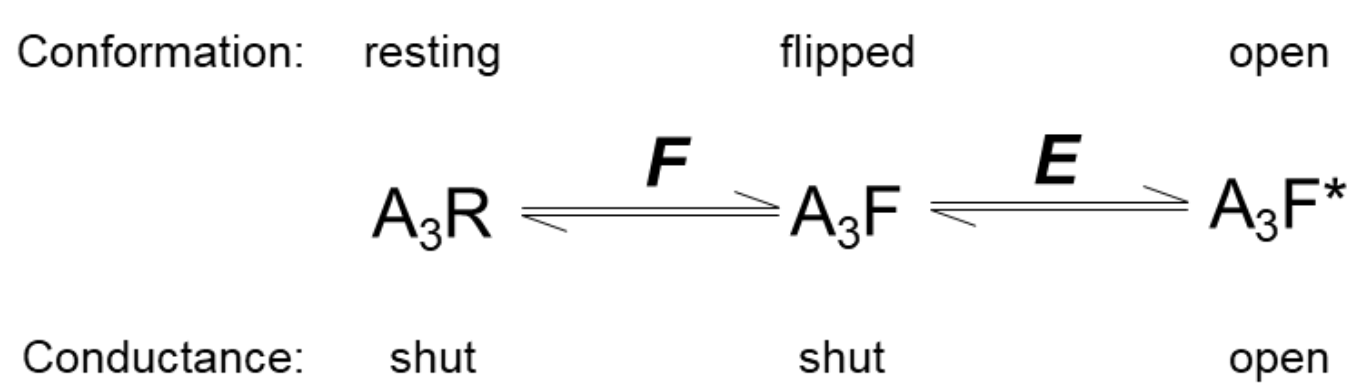

Figure 3 The case of a glycine receptor in the presence of a saturating concentration of agonist, so all three binding sites are fully occupied. The equilibrium constant for the isomerisation between resting ($A_3R$) and flipped ($A_3F$) conformations is denoted as *F*, and the equilibrium constant for the isomerisation between flipped and open ($A_3F^*$) conformations is denoted as ***E*.**

In this case, the maximum response corresponds to the maximum fraction of receptors in the open $A_3F^*$ state, $p_{\mathrm{open}}^{\max}$. This is determined by both of the equilibrium constants in Figure 3. It can be written as

$$p_{\mathrm{open}}^{\max} = \frac{EF}{1 + F + EF} = \frac{E_{\mathrm{eff}}}{1 + E_{\mathrm{eff}}}$$

where we have defined an effective efficacy as

$$E_{\mathrm{eff}} \equiv E\,\frac{F}{1+F}$$

For a single receptor channel, the response can, under favourable conditions, be measured directly from experimental recordings as the fraction of time spent in open state. This is the only case in which the response to an agonist can be measured on an absolute scale, rather than being measured relative to an arbitrary maximum.

Ever since 1957 [6], it had been assumed that partial agonists on ion channels were partial because the opening reaction was inefficient ($E$ is small). But it was proposed by Lape *et al*, 2008 [14], that partial agonists are partial because the first step is inefficient ($F$ is small), but the second step, the shut-open transition, is much the same for both partial and full agonists. This was later corroborated by Mukhtasimova *et al* (2009) [26].

## 5. The case of G-protein-coupled receptors

Given their great importance to both physiology and disease, the structure and function of G-protein-coupled receptors (GPCRs) has been at the centre of pharmacological research for decades. However, the transduction mechanisms for GPCRs are too complex to allow an experimenter to measure separately the affinity and efficacy of an agonist. The 'cubic ternary complex' (CTC) mechanism, which has been used to describe the interaction between agonists, GPCRs and G-proteins, contains seven free equilibrium constants in its simplest form. Six of the seven equilibrium constants reflect various aspects of efficacy; only one provides information about agonist affinity and, hence, the agonist binding site. This makes structure-activity relationships for agonists of GPCRs difficult to interpret [3]. The identification of the agonist binding site is also not straightforward, though structural and molecular dynamics studies can certainly help.

Responses to GPCR agonists that can be readily measured are often far removed from agonist binding, and several steps separate agonist binding and receptor activation from the response being measured (e.g., the accumulation of cyclic AMP). Even if one were to measure more proximal events in the transduction pathway, such as the association of a G-protein with a GPCR in response to agonist application, there are still too many steps, and too many unknowns, to make any meaningful conclusions about the affinity and efficacy of an agonist. For example, the equilibrium between the G-protein and the GPCR must be defined, and it is not even possible to define the G-protein concentration, insofar as G-proteins are membrane delimited, let alone account for the depletion of free G-protein as it binds to the receptor [27].

This is plain to see if one considers the maximum response elicited by an agonist in the simple CTC mechanism, that is, the maximum fraction of receptors in the agonist- and G-protein-bound active state in the presence of a saturating agonist concentration

$$p_{active}^{max} = \frac{E_{\mathrm{G}} c_{\mathrm{G}}}{1 + E_{\mathrm{A}} + c_{\mathrm{G}} + E_{\mathrm{G}} c_{\mathrm{G}}}$$

where we define

$$c_{\mathrm{G}} \equiv \frac{[G]}{K_{\mathrm{G}}}$$

The maximal response depends on two efficacy terms: $E_{\mathrm{A}}$, the equilibrium constant for the isomerisation of the receptor between its inactive and active state once the agonist has bound (as in section 2); and $E_{\mathrm{G}}$, the equilibrium constant for the isomerisation of the receptor between its inactive and active state following the binding of agonist and G-protein. The maximal response also depends on [G], the concentration of G-protein, and $K_G$, the dissociation constant for the binding of G-protein to the agonist-bound inactive state of the receptor.

Measurements of the maximal response to an agonist in these types of experiment cannot provide the agonist's efficacy at the level of the receptor; rather, these measurements provide information about the overall 'coupling efficiency' of the pathway being studied. Consequently, it is possible for an agonist which is full at the receptor level to appear partial at the cellular level, and vice versa. The agonist concentration required to elicit the half-maximal response (or half-maximal binding) is dependent on all seven free equilibrium constants in the CTC mechanism and so cannot be easily interpreted either. These experiments provide useful information about the cellular responses evoked by different agonists under different conditions but provide little information about the properties of the agonist-receptor interaction.

A great deal of our understanding of GPCR function comes from experiments which measure events downstream of receptor activation. However, an understanding of an agonist's properties will require knowledge of the conformational changes of the receptor itself, as is the case for ligand-gated ion channels. There are currently limited methods for directly observing agonist-induced changes in the conformation of a GPCR in real time. One way to do this is to use single-molecule intramolecular resonance energy transfer measurements. Such measurements may be difficult to interpret because they depend on the relative locations of the donor and acceptor (chosen by the experimenter). How the distance between a given donor-acceptor pair relates to the various states of the receptor may not be clear as much of our understanding of agonist-induced conformational changes in GPCRs comes from static crystallographic and spectroscopic experiments. It also remains to be seen whether such experiments have the resolution required to yield estimates of the rate constants in a reaction mechanism.

Attempts have been made to fit the rate constants of realistic mechanisms of GPCR function to experimental data [28,29], but the fits are far too over-parameterised to provide good estimates.

## 6. Conclusions

Stephenson greatly advanced pharmacology when he recognised that it's impossible to describe the action of an agonist at equilibrium with a single number; his addition of one more parameter – efficacy – provided a qualitative description of partial agonism. The simple example in section 2 clearly shows that Stephenson's framework does not allow the separate estimation of values of affinity and efficacy. These two properties of agonists are inextricably linked, and, in most sorts of experiment, they cannot be isolated from one another. Thus far, only single-channel recording of ligand-gated ion channel currents has allowed the separation of affinity and efficacy. Agonist binding experiments and structure-activity relationships are of limited use because the binding of, and response to, an agonist is influenced by *both* its affinity and efficacy.

Stephenson's mistake remained unnoticed until 1987 [2], and is still often misunderstood even now. The original, erroneous framework still pervades pharmacology and has been propagated in lecture theatres, textbooks and the scientific literature for decades. Furthermore, more recent models of agonist action, such as those proposed by Furchgott (1966) [30] and Black and Leff (1983) [31], incorporated the same framework used by Stephenson, and so they too cannot separate affinity and efficacy. Although the classical equations can, under limited conditions, take the same form as those that describe realistic mechanisms, it's agreed that equilibrium measurements can't estimate the quantities with physical meaning – the underlying equilibrium constants [32].

Generations of students of biochemistry have been expected to understand the Michaelis-Menten mechanism, yet it is still rare for students of pharmacology to be taught its equivalent, the del Castillo-Katz mechanism. Indeed, the affinity-efficacy problem is not included in the core curricula for pharmacology undergraduates proposed by the British Pharmacological Society, or by IUPHAR [33].

We believe that the affinity-efficacy problem should be part of the core curriculum for students who study pharmacology as a science.